\documentclass[iop]{emulateapj}
\usepackage{hyperref}

\slugcomment{Submitted, The WSPC Handbook of Astronomical Instrumentation: Volume 3: UV, Optical \& IR Instrumentation: Part 2}

\shorttitle{Automated Adaptive Optics}
\shortauthors{Baranec et al.}

\begin{document}


\title{Automated Adaptive Optics}


\author{Christoph Baranec\altaffilmark{1}, Reed Riddle\altaffilmark{2}, and Nicholas M. Law\altaffilmark{3}}
\email{baranec@hawaii.edu}


\altaffiltext{1}{Institute for Astronomy, University of Hawai`i at M\={a}noa, Hilo, HI 96720-2700, USA}
\altaffiltext{2}{Division of Physics, Mathematics, and Astronomy, California Institute of Technology, Pasadena, CA 91125, USA}
\altaffiltext{3}{Department of Physics and Astronomy, University of North Carolina at Chapel Hill, Chapel Hill, NC 27599-3255, USA}

\begin{abstract}

Large area surveys will dominate the forthcoming decades of astronomy and their success requires characterizing thousands of discoveries through additional observations at higher spatial or spectral resolution, and at complementary cadences or periods. Only the full automation of adaptive optics systems will enable high-acuity, high-sensitivity follow-up observations of several tens of thousands of these objects per year, maximizing on-sky time. Automation will also enable rapid response to target-of-opportunity events within minutes, minimizing the time between discovery and characterization. 

In June 2012, we demonstrated the first fully automated operation of an astronomical adaptive optics system by observing 125 objects in succession with the Robo-AO system. Efficiency has increased ever since, with a typical night comprising 200-250 automated observations at the visible diffraction limit. By observing tens of thousands of targets in the largest-ever adaptive-optics surveys, Robo-AO has demonstrated the ability to address the follow-up needs of current and future large astronomical surveys.

\end{abstract}



\section{The need for efficient adaptive optics}\label{aao_sec1}

In 1998, J. Hardy identified several future directions for adaptive optics (AO) including a ``Simplified operator interface with the adaptive optics, the goal being to make the operations mostly automatic ... including control of the beacon laser" \citep{Hardy}.  It is now, during the current era of extremely large astronomical surveys, that there is a clear need for automated AO imaging and spectroscopy to fully characterize the large numbers of discoveries \citep{Kulkarni12}. 

For example, NASA's \textit{Kepler} mission identified more than 3,000 stars with repeating transit signals indicative of transiting exoplanets. Follow-up AO imaging of these objects is used to determine the sources contributing to the \textit{Kepler} light curves in order to rule out astrophysical false positives, to accurately measure the radii of the detected planets with respect to their host star, to measure the effect of stellar multiplicity on exoplanet formation, and to determine the physical association of detected stellar companions \citep{Morton11, Kraus16, Law14, Baranec16, Ziegler17, Atkinson17}. Upcoming transit missions such as NASA's Transiting Exoplanet Survey Satellite \citep{Ricker15} and ESA's PLAnetary Transits and Oscillations \citep{Rauer16} are expected to discover many more exoplanet systems than \textit{Kepler}. 

General wide field surveys such as the Zwicky Transient Facility \citep{Terziev13}, Evryscope \citep{Evr15}, Pan-STARRS \citep{Deacon17} and ESA’s Gaia \citep{dB15} are detecting hundreds of thousands of close stellar systems, partially resolved in imaging data, visible from eclipses and other time-domain phenomena, or forming confusing asterisms in crowded fields. For many science goals, each survey requires high-angular-resolution follow-up to confirm systems or remove confusing effects from the crowded targets.   Only very efficient AO can confirm and measure the properties of a significant fraction of these close stellar systems.

Outside our galaxy, adaptive optics images are also crucial in disentangling transient events from their environments: determining whether they are hosted in a faint galaxy, are associated with precursor images, or if their light curves are contaminated by nearby sources \citep{Ofek07, Li11, Cao13}. 

These science objectives can be loosely classified into three major categories: large population studies, rapid target characterization, and long-term monitoring. Table~\ref{aao_tab1} highlights many areas where automated AO is necessary to enable science that was previously thought to be impractical with exiting facilities.

\begin{center}
\begin{table*}
\caption{Example automated AO science}
\centering
{\begin{tabular}{@{}ll@{}} \toprule
Science topic & Number of targets and/or cadence \\
\colrule
Transiting exoplanet hosts & $>$10,000 targets \\
Wide exoplanets and brown dwarfs & 5,000 -- 10,000 targets \\
Large survey stellar multiplicity & $>$10,000 targets \\
Transient characterization & Rapid response, declining follow-up cadence \\
Astrometric microlensing & Dozens of high-cadence events per year \\
Solar sys. small body nucleus characterization, & Few night response, $\sim$10 Manx, Centaurs, \\
exopause searches, surface minerology & and comets per year \\
Discover/monitor lensed quasars & $>$25,000 + monitoring 3 nights/mo. \\
Monitoring planetary weather & Snapshot 2--3 times per night \\
Monitoring jets, outflows and shocks & Several times per year \\
\botrule
\end{tabular}
}
\label{aao_tab1}
\end{table*}
\end{center}

\section{General philosophy for automation}

Although the scientific community is currently well served by flexible stellar/laser or boutique high-contrast AO systems for limited numbers of targets, their scarcity, complexity, and competition for observing time on large apertures limits their suitability for following up large numbers of targets \citep{Hart10a, Davies12}. The key for automated AO is to focus primarily on those scientific objectives that uniquely require high-time-efficiency observing.

The overarching philosophy in the automation of adaptive optics has been to prioritize reliability, predictability, and ease-of-use over more traditional AO metrics such as Strehl ratio, achievable contrast or sky coverage. While the requirements on the latter should be well understood and meet the science need, they should not be the driving factors in system design because the primary metric is simply the number of targets that can be observed at high-angular-resolution.
 
\textbf{Reliability}:
The entire observing system, i.e., telescope, AO, science instruments, control software, and data reduction pipeline, needs to be very reliable in operation. All hardware components should be sufficiently past their beta-testing phase of development and be easily spared or serviceable if necessary. Software should be designed to be as modular as possible in order to increase reliability and delineate areas of complexity.  It should be straightforward to identify and recover from error states during operations to minimize or eliminate the need for human intervention. During the commissioning phase of the instrument, it is necessary to exercise and stress the system on sky, identify failure modes, fix errors, and repeat the process until the robotic system is capable of independent operation. 
 
\textbf{Predictability}:
The operation of the system should be designed intentionally such that the observing system follows logical paths. The output of the system should be repeatable and deterministic given a set of input conditions and parameters such as seeing, wind speed, Zenith angle, etc. Effort should be placed upon minimizing the effects of conditions where one has some control, e.g., air conditioning and effective dome venting to ensure thermal equalization of the telescope and mitigation of dome seeing. Diagnostics should be in place to measure extraneous factors which serve as predictors for the quality of an observation, oftentimes as criteria for suitable observing conditions or determining the need for repetition of a particular observation.

\textbf{Ease-of-use}:
The observing system should operate itself with no or minimal human oversight, as opposed to a ``push-button'' AO system that can be driven by non-experts (e.g., at most major observatories where a dedicated telescope operator or support scientist operates the AO system). Any time an operator needs to make a decision, precious time is lost. Scientists should limit their interaction with the system to adding observation requests and retrieving their data. 

By following these guiding priorities, is it clear that it is likely prohibitively difficult to develop an immensely flexible AO platform that can support a suite of different automated modes for a range of seeing conditions and multiple target types. Instead, limiting the modes of operation minimizes the number of steps and variables needed to perform an observation, containing the development costs and schedule of an automated system. This has the natural consequence of being favorable to queue type observing lists and simplifying requirements on data reduction and analysis pipelines. Focusing the development on creating a reliable robotic system that can complete the science mission builds a robust and capable system.

%
%
%
%
%
%
%

\section{Automation enabled by technological maturity}

While scientific requirements drive the need to automate AO systems, technological maturity limits their development. Adaptive optics systems are almost exclusively developed as prototype instruments that take advantage of a new architecture and/or component development. Projects are then required to beta-test a subset of new technology while primarily building on established methods and technology, and often times only the largest aperture observatories have sufficient funding for this development. To develop automation as the next technological leap, the component technologies need to be proven as reliable and not be the primary cost drivers.

In the decades since the first scientific adaptive optics systems were put into use, many component technologies have continued to improve in functionality, reliability and cost. A non-exhaustive list of examples of many relevant innovations for automated AO follows.
\begin{itemize}
\item Optical phase correctors: The Visible Light Laser Guidestar Experiments (ViLLaGEs) at Lick Observatory \citep{Gavel2008} pioneered the use of Microelectromechanical systems (MEMS) deformable mirrors (DMs) in astronomical adaptive optics. MEMS with hundreds of actuators and several wavelengths of phase correction are now available as off-the-shelf catalog items at an order of magnitude cost less than traditional piezo-stack deformable mirrors. So long as the mirrors are sealed against humidity, they have proven to be extremely reliable, have negligible hysteresis for most applications and have no required feedback on the commanded position of the individual actuators \citep{Bifano2008}.

\item Laser guide stars: Sodium excitation dye and sum frequency solid-state laser guide stars are relatively expensive due to non-existent commercial applications and require extensive maintenance schedules for reliable operation due to their prototypical nature. While comparable in cost, new Raman fiber amplifier based 589 nm lasers promise to improve upon both the photon return and reliability and have begun use at large telescopes in 2016. In contrast, artificial guide stars created by Rayleigh scattering laser light off of air molecules can be driven by turn-key Q-switched lasers developed for the Silicon wafer processing industry\footnote{The photon return per Watt is maximized around $\lambda\sim450$ nm with a $\sim25\%$ drop off at the doubled and tripled Nd:YAG wavelengths of $\lambda =$ 355 and 532 nm \citep{Georges2003}. Pulse rates need to fall between the minimum AO loop rate and the beacon pulse round-trip time, e.g., $\sim$1 kHz to 15 kHz for a beacon at 10km.} (with typical mean times between failures exceeding 10,000 hours). Additionally, Rayleigh laser beacons at wavelengths below the threshold of human vision ($\lambda<$400 nm) can be granted a waiver from Federal Aviation Administration control measures (e.g., human spotters, transponder detectors) because of their inability to flash blind pilots or produce damaging levels of radiation during brief exposures for typical, $<$100 W, power levels.

\item Wavefront reconstructor controllers: Early generations of adaptive optics systems relied on either analog or very specialized digital computing technologies for wavefront reconstruction and control. As of a decade ago, personal computers (PCs) have proven themselves capable of running modest actuator count infrared AO systems on large telescopes  \citep{Velur2006, Vaitheeswaran2008}, and more recently are able to drive exoplanet AO systems \citep{Dekany2013}. PCs support many common software languages that have been used to code wavefront reconstructor controller routines, e.g., C, C++, G, MATLAB and Python, lowering the barrier to new development or to making modifications to existing code.

\item Science cameras: A major complication with observing astrophysical objects with AO is that their peak brightness will not be known until the actual observation. Often one relies on seeing limited or poorly sampled catalogs to plan observations which may not be accurate and variable atmospheric conditions result in a constantly changing Strehl ratio. Long integrations with a conventional CCD or infrared array detector may need to be repeated if there is insufficient radiant flux in a single image (additional images lead to more accumulated read-noise) or if the object saturates or exceeds the linear dynamic range of the detector. With the advent of fast readout, sub-electron readout detectors in astronomy (e.g., electron multiplying (EM) CCDs \citep{Tubbs2002, Law06b} and infrared avalanche photodiode arrays \citep{Baranec2015}), the effects of inaccurate or variable peak brightness are significantly mitigated. Beyond additional data storage requirements, there is minimal noise penalty for taking a long series of short exposures, each of which are able to span the entirety of the dynamic range of the detector. An additional benefit of having a series of short exposures is the ability to register each detector read to compensate for any uncorrected stellar displacement, or to apply other post-processing techniques such as speckle interferometry. 

\item Host telescopes: Full automation of the host telescope, as has been done with other astronomical survey facilities \citep{Brown2013, Vogt2014}, is crucial to maintaining efficient observations. Activities previously handled by telescope operators and night assistants, such as environmental safety, pointing of the telescope and instrument calibrations, are controlled and sequenced by a computer. While automated telescopes today are almost exclusively modest in aperture, no physical limitation prohibits the full automation of large telescope apertures\footnote{The Large Synoptic Survey Telescope will soon be the largest automated astronomical telescope in existence. }.

\end{itemize}

\section{CAMERA}

The idea for an autonomous laser guide star AO system was first proposed by M. Britton, \textit{et al.} \citep{Britton2008} in 2007. The \underline{C}ompact \underline{A}utonomous \underline{ME}MS-based \underline{R}ayleigh \underline{A}O (CAMERA) system was intended to be a low-cost AO system mounted to a small robotic telescope to perform survey astronomy in queue-mode as well as monitor VOEvent feeds to characterize important transient objects. A dichroic would split the final AO output to visible and near infrared cameras: one of the two would be used as tip-tilt sensor driving a fast-steering mirror to correct stellar image displacement while the other camera would be used to capture science images. 

Inspiration for the project stemmed from several developments in AO engineering: the development of PC reconstructors and economical wavefront sensors as part of the multiple guide star tomography demonstration at Palomar Observatory \citep{Velur2006}; the ongoing development of cost effective MEMS DMs at Lick Observatory; the use of industrial lasers at UV wavelengths at Mt. Wilson \citep{Thompson02} and for multiple Rayleigh beacons at the MMT \citep{Hart10b}; and the availability of the recently roboticized 1.5-m telescope at Palomar \citep{Cenko06}.

A proof-of-concept laboratory AO system was then developed at Caltech that used a Boston Micromachines Multi-DM, a PI fast steering mirror, a PC reconstructor and a Shack-Hartmann wavefront sensor that used a SciMeasure camera with an E2V CCD39 detector. It achieved a closed-loop update rate of 100 Hz, corrected wavefront aberrations induced by a spinning disk of plastic, and could be controlled through a simple web-based interface.

\section{Robo-AO}

To transform the CAMERA concept and laboratory system to a scientifically competitive on-sky AO system, we initiated the Robo-AO project \citep{Baranec2013, Baranec2014} in 2009 as a collaboration between Caltech Optical Observatories and the Inter-University Centre for Astronomy and Astrophysics (IUCAA). Our first objective was to build and demonstrate a Robo-AO system for the Palomar 1.5-m telescope, then clone the system and install on the IUCAA Girawali Observatory 2-m telescope. While we nominally followed the CAMERA blueprint, we made two necessary major deviations. We postponed the development of the infrared science camera due to its significant relative cost and potentially lower scientific impact compared to the visible science camera. We also changed the control software architecture from an all-encompassing monolithic program to a set of modular sub-systems that would be overseen by a master scheduler and watch-dog processes. A detailed description of the first as-built Robo-AO system follows.

\subsection{Hardware}

\begin{figure}[hb]
\centerline{\includegraphics[width=3.25in]{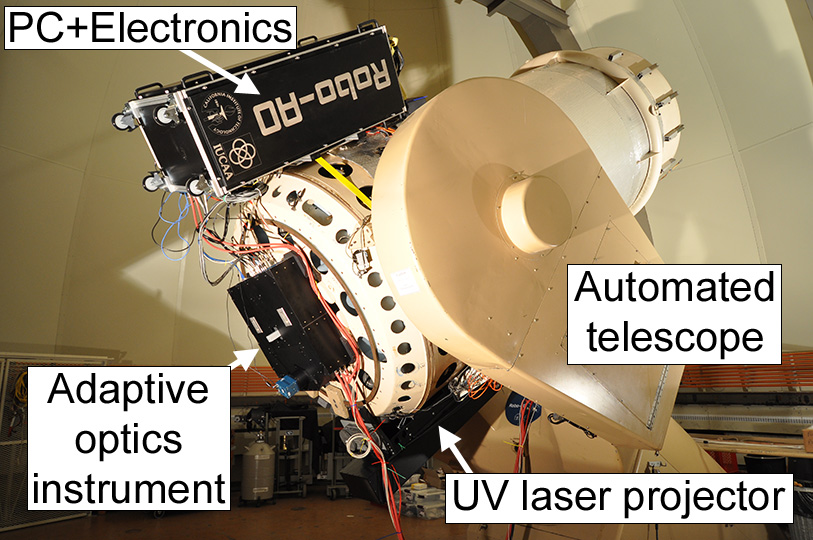}}
\caption{The Robo-AO system on the automated 1.5-m telescope at Palomar Observatory.}
\label{fig_palomar}
\end{figure}

Robo-AO comprises several main systems: the UV laser projector; an instrument mounted at the Cassegrain focus of the telescope (Fig. \ref{fig_palomar}), and a set of electronics including the PC reconstructor and control computer.


\begin{figure}[hb]
\centerline{\includegraphics[width=3.25in]{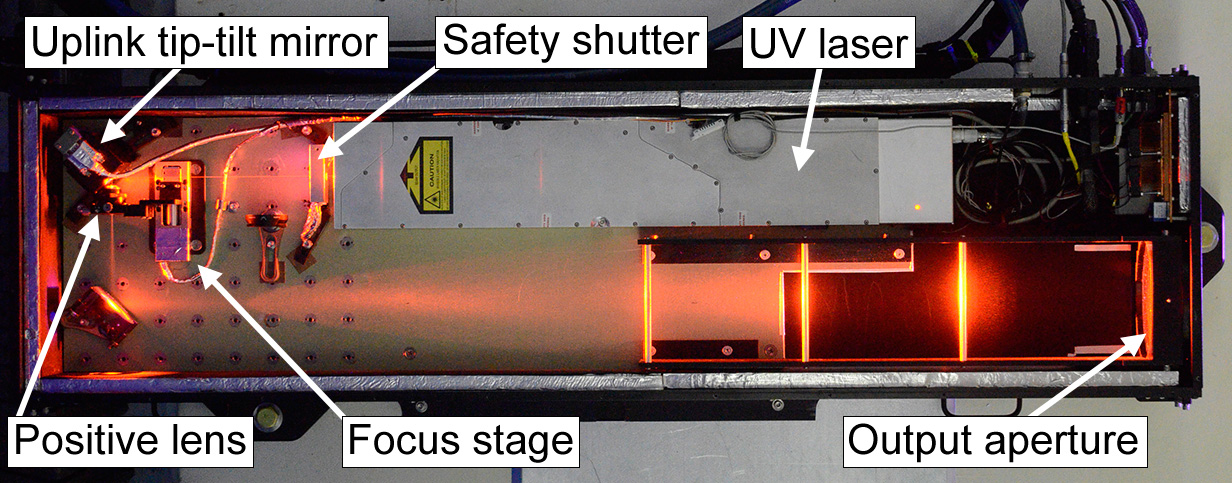}}
\caption{The inside of the Robo-AO laser projector.}
\label{fig_laser}
\end{figure}

The UV laser projector enclosure is 1.5 m $\times$ 0.4 m $\times$ 0.25 m $\times$ $\sim$70 kg, and attaches to the side of the 1.5-m telescope (Fig. \ref{fig_laser}). Inside are a commercial pulsed 12-W ultraviolet laser (35 ns pulses every 100 $\mathrm{\mu}$s, $\lambda$=355nm); a redundant safety shutter; and an uplink tip-tilt mirror to both stabilize the apparent laser beam position on sky and to correct for up to 2$'$ of differential pointing errors. A positive lens on an adjustable focus stage expands the laser beam to fill a 15 cm output aperture lens that is optically conjugate to the tip-tilt mirror. The laser beam is coaligned with the bore-sight of the principal telescope with its waist focused to a 10 km line-of-sight distance.

\begin{figure}[ht]
\centerline{\includegraphics[width=3.25in]{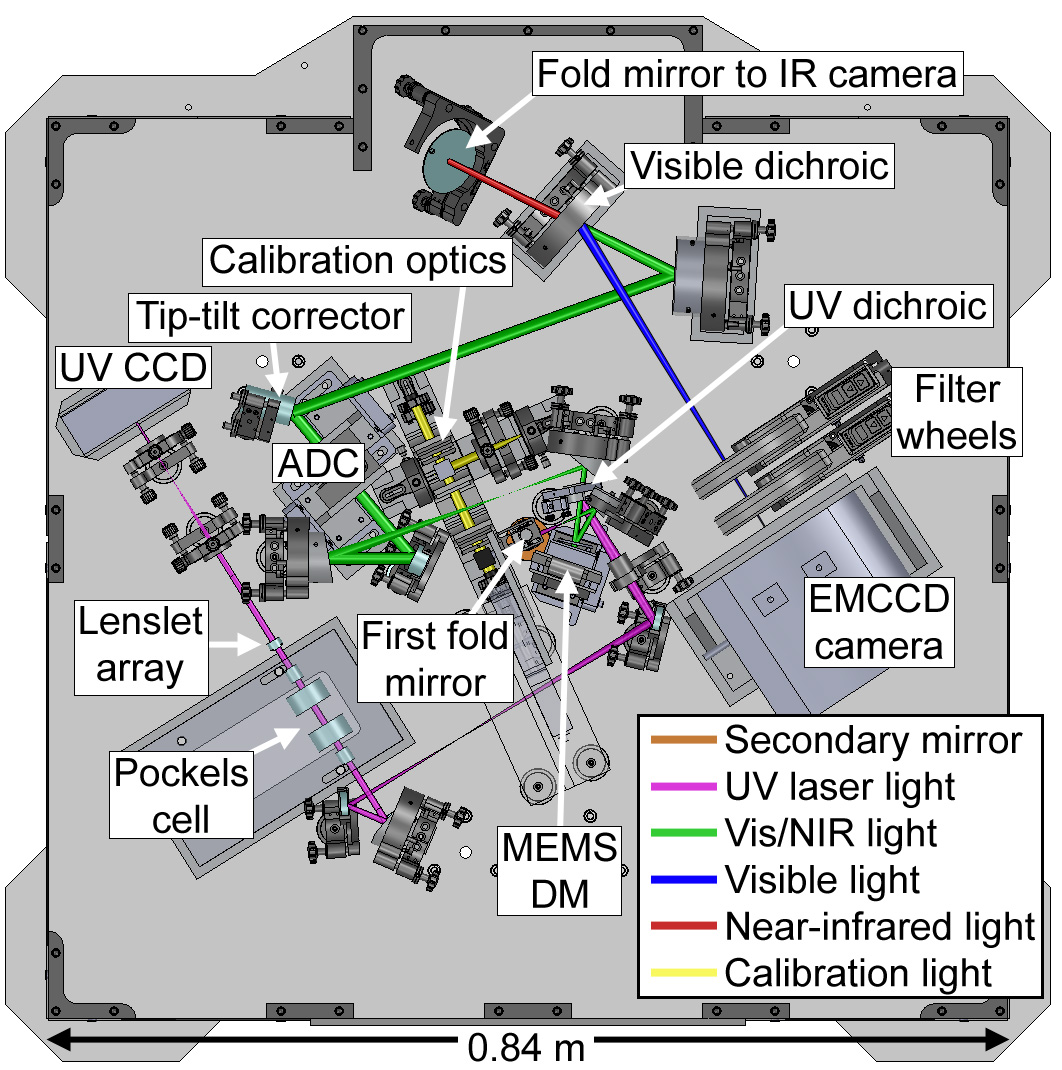}}
\caption{The inside of the Robo-AO Cassegrain adaptive optics system.}
\label{fig_cass}
\end{figure}

The adaptive optics system and science cameras reside within a Cassegrain mounted structure of approximate dimensions: 1 m $\times$ 1 m $\times$ 0.2 m (Fig. \ref{fig_cass}). Light from the telescope secondary mirror enters the instrument and is intercepted by a fold mirror which directs a 2$'$ diameter field to a dual off-axis parabolic (OAP) mirror relay. The first fold mirror is on a linear motorized stage that can be moved out of the beam path, revealing internal calibration optics that simultaneously simulate the ultraviolet laser focus at 10 km and an optical/infrared incandescent source at infinity, matching the telescope focal ratio and exit pupil position.
The first OAP images the telescope pupil onto a 12 $\times$ 12 actuator MEMS DM. 

After reflection off the DM, the UV laser light is selected off with a UV dichroic mirror and refocused to a 5$''$ field stop. The light is collimated with a reversed OAP to cancel out coma from the first OAP. The collimated light passes through a dual-crystal $\beta$-BaB$_2$O$_4$ Pockels cell between two crossed linear polarizing cubes. The Pockels cell is used as a high speed shutter to limit the backscattered Rayleigh laser light to a range of 450 m around the 10 km beam waist. An 11 $\times$ 11 lenslet array is located at a pupil and the Shack-Hartmann pattern is demagnified onto an E2V UV optimized CCD39 detector (80$\times$80 pixels; 72\% quantum efficiency at $\lambda$ = 350 nm). The pixels are binned by a factor of 3 and the slope of each subaperture is calculated from 2 $\times$ 2 binned pixels. The AO control loop operates at the 1.2 kHz frame rate of the detector, with an effective control bandwidth of 90--100 Hz.

The visible and infrared light passes through the UV dichroic and is refocused by another OAP. The light is then relayed by a second OAP relay which includes a tip-tilt corrector and an atmospheric dispersion corrector (ADC; 400 nm $< \lambda <$ 1.8 $\mu$m). The final relay element creates a telecentric F/41 beam that is split by a visible dichroic mirror at $\lambda$ = 950 nm with the infrared light directed via a fold mirror to an external camera port. 

The visible light is captured by an EMCCD (E2V CCD201–20) camera with a 44$''$ square field of view and 0.043$''$ pixel scale. Frames are continually read at a rate of 8.6 Hz during science observations, allowing image displacement, that cannot be measured using the laser system \citep{Rigaut92}, to be removed in software based on the position of a $m_V \leq 16$ guide star within the field of view. The EM gain is set before observations based on the target's magnitude to minimize read-noise while leaving sufficient dynamic range if the targets are brighter than anticipated. Typically, an EM gain of 300 (readnoise $<$ 0.2 e$^{-}$) is appropriate for targets $m_V > 13$, and is incrementally decreased to 25 (readnoise = 1.9 e$^{-}$) for targets as bright as $m_V = 2$.   

\newpage

\subsection{Software}

The Robo-AO master robotic sequencer software \citep{Riddle12} controls the telescope, adaptive optics system, laser, filter wheels, and science camera; executing all operations that otherwise would have been performed manually, allowing greatly improved observing efficiency. The software to control each hardware subsystem was developed as a set of individual modules in C++, and small standalone test programs have been created to test each of the hardware interfaces. This modular design allows the individual subsystems to be stacked together into larger modules, which can then be managed by the master robotic sequencer.

The execution of an observation starts with a query to an intelligent queue scheduling program \citep{Riddle14} that selects a target. The robotic sequencer will then point the telescope, while simultaneously selecting the appropriate optical filter and configuring the science camera, laser and adaptive optics system. A laser acquisition process to compensate for differential pointing between the telescope and laser projector optical axes, caused by changing gravity vectors, begins once the telescope has completed pointing at the new target. A search algorithm acquires the laser by moving the uplink steering mirror in an outward spiral pattern from center until 80\% of the wavefront sensor subapertures have met a flux threshold of 75\% of the typical laser return flux. Simultaneous with the laser acquisition process, the science camera is read out for 10-20 s with no adaptive optics compensation and with the deformable mirror fixed to obtain a contemporaneous estimate of the seeing conditions through the telescope. For targets identified in the queue brighter than $m_V = 12$, the telescope is offset to center the star on the detector; for fainter stars the robotic system will only recenter if the brightest star is within the center quarter frame, thus avoiding centering on a bright companion star near the edge of the frame and pushing the fainter science star out. Upon completion of laser acquisition, a new wavefront sensor background image is taken, the adaptive optics correction is started and an observation with the science camera begins.

During an observation, telemetry from the adaptive optics loop is used to maintain telescope focus and detect significant drops in laser return flux. Slow drifts in the focus mode of the deformable mirror are measured and offloaded to the secondary mirror to preserve the dynamic range of the deformable mirror. Focus on the deformable mirror is measured by projecting the commanded actuator values to a model Zernike focus mode. A median of the last 30 focus values, measured at 1-s intervals, is calculated. If the magnitude of this value exceeds 220 nm peak-to-valley surface of focus on the deformable mirror, equivalent to a displacement of 20 $\mathrm{\mu}$m of the Palomar 1.5-m telescope secondary mirror, then the secondary is commanded to change focus to null out this value. Focus corrections may not be applied more than once every 30 s and are restricted to less than 50 $\mu$m of total secondary motion to avoid runaway focus. The laser return flux is also measured at simultaneous 1-s intervals. If the laser return drops below 50 photoelectrons per subaperture on the wavefront sensor for more than 10\% of the values used to calculate the median focus, e.g., due to low-altitude clouds or extremely poor seeing (greater than 2.5$''$), any focus correction is ignored due to the low certainty of the measurement. Additionally, if the return stays below 50 photoelectrons per subaperture for five consecutive seconds, the observation is immediately aborted, the target is marked as ``attempted but not observed'' in the queue, and a new target is selected for observation.

The intelligent queue is able to pick from all targets in a directory structure organized by scientific program, with observation parameters defined within Extensible Markup Language (.XML) files. Users are able to load targets directly, or through text file interpreters or a web management system. The queue uses an optimization routine based on scientific priority, slew time, telescope limits, prior observing attempts, and laser-satellite avoidance windows to determine the next target to observe. In coordination with US Strategic Command (USSC), we have implemented measures to avoid laser illumination of satellites. To facilitate rapid follow-up observations we have developed new de-confliction procedures which utilize the existing USSC protocols to open the majority of the overhead sky for possible observation without requiring preplanning. By requesting predictive avoidance authorization for individual fixed azimuth and elevation ranges, as opposed to individual sidereal targets, Robo-AO has the unique capability to undertake laser observations of the majority of overhead targets at any given time.

A system monitor manages the information flow of the status of the individual subsystems that comprise the entire robotic system. It detects when one of the software subsystems has an error, crashes, or other issues that
might hinder the proper operation of the system. Errors are logged and the operation of the automated
observations are stopped until the subsystem daemon can clear the issue. If the subsystem cannot correct the error,
the automation system can take steps, up to and including restarting subsystems, in an attempt to continue
operations. If it is unable to restart the system, it shuts everything down, leaving the system in a safe state.

\subsection{Data reduction and analysis}

Upon completion of an observation, the data are compressed and archived to a separate computer system where the data are immediately processed. A data reduction pipeline \citep{Law14spie} corrects each of the recorded frames for detector bias and flat-fielding effects, and automatically measures the location of the guide star in each frame. The region around the star is up-sampled by a factor of four using a cubic interpolation, and the resulting image is cross-correlated with a diffraction-limited point spread function for that wavelength. The frame is then shifted to align the position of greatest correlation to that of the other frames in the observation, and the stack of frames is coadded using the Drizzle algorithm to produce a final high-resolution output image sampled at twice the resolution of the input images. 
For science programs which require the detection and contrast ratio measurement of closely separated objects, an additional point-spread-function (PSF) subtraction and analysis pipeline can be started upon completion of the data reduction pipeline. This pipeline distinguishes astrophysical objects from residual atmospheric and instrumental wavefront errors and corresponding speckles in the image plane using a modified Locally Optimized Combination of Images  \citep{L07b} algorithm. The algorithm selects a combination of similar PSFs from the hundreds of other observations of similar targets during that night. The combination of PSFs is used to create a model PSF which is then subtracted from each image and potential companions are flagged.

\begin{figure}[t]
\centerline{\includegraphics[width=3.25in]{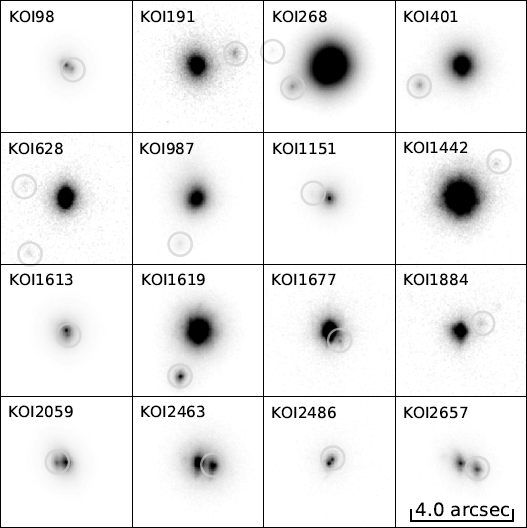}}
\caption{Example Robo-AO images of \textit{Kepler} candidate exoplanet host stars with stellar blends that contribute additional light to the measured light curves.}
\label{ra_panel}
\end{figure}

The rate at which final images are processed lags only slightly behind the data capture rate: a full night's set of data is typically finished before the next night of observing. Example cutouts of fully reduced Robo-AO images showing multiple stars appears in Figure \ref{ra_panel}. Data are available as full-frame, cutout and PSF subtracted images with estimates of achievable contrast vs. separation, and can be downloaded over scp or through the use of wget scripts. 

\section{Automated AO results}


On August 14th, 2011, we closed the the high-order AO loop on Robo-AO for the first time, achieving a clear diffraction-limited core in the fast frame rate visible camera images. Over the following year, we improved the operation of the system with a focus on the master robotic sequencer software. On June 19th, 2012, Robo-AO demonstrated fully automated operation by observing 125 objects in succession with no human assistance. Initially, AO setup overhead times were on the order of 60s (excluding telescope slew time). With further software optimization, this was reduced to $\sim$40s in 2014; and, with a change in the power switching module that controls the operation of Pockels cell shutter, this is currently around $\sim$20s.  With small slews (i.e. a few degrees), Robo-AO achieved total overhead times (including telescope slew) of less than 40s on the Palomar 1.5m telescope by 2015.

\begin{figure}[hb]
\centerline{\includegraphics[width=3.25in]{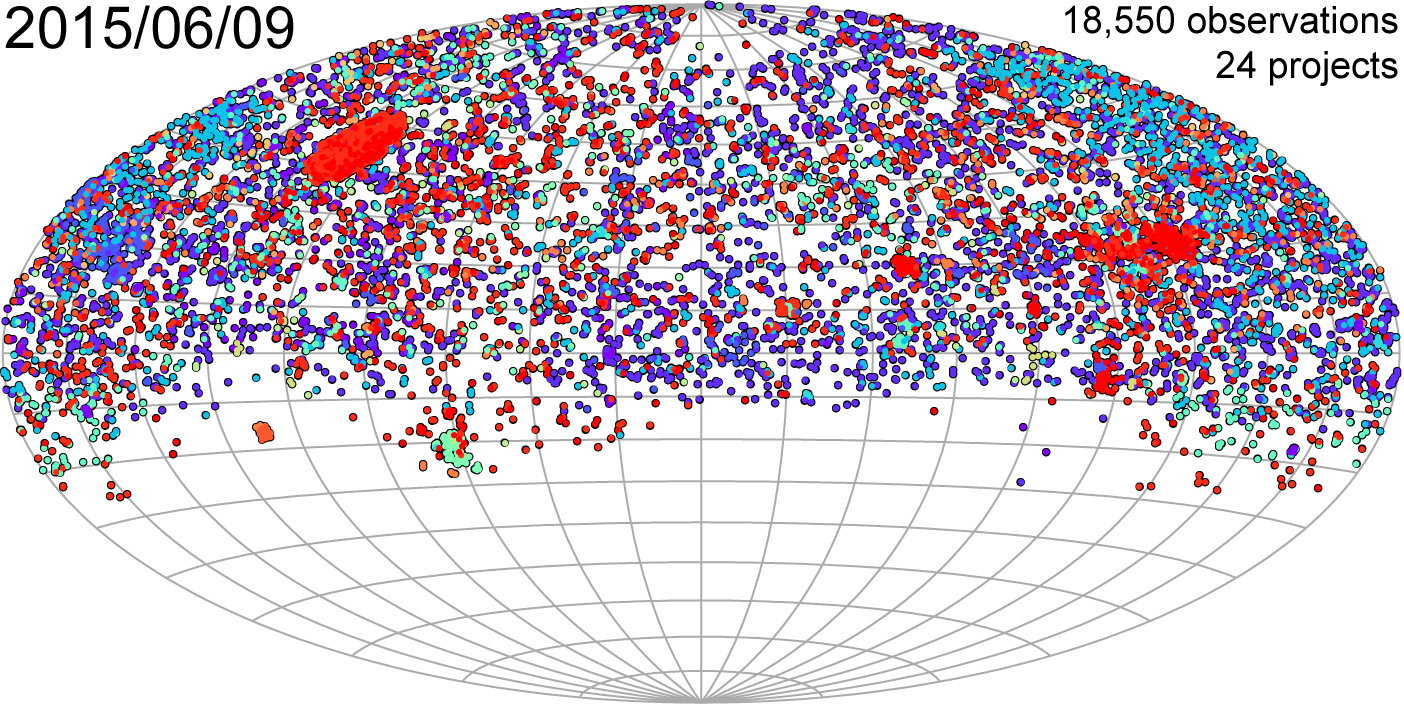}}
\caption{High-angular-resolution visible-light AO imaging performed by Robo-AO. Each of the 18,550 points on this graph is an observation performed by the robot and automatically processed by the pipeline into a final science-quality image.}
\label{sky_map}
\end{figure}

\begin{figure}[hb]
\centerline{\includegraphics[width=3.25in]{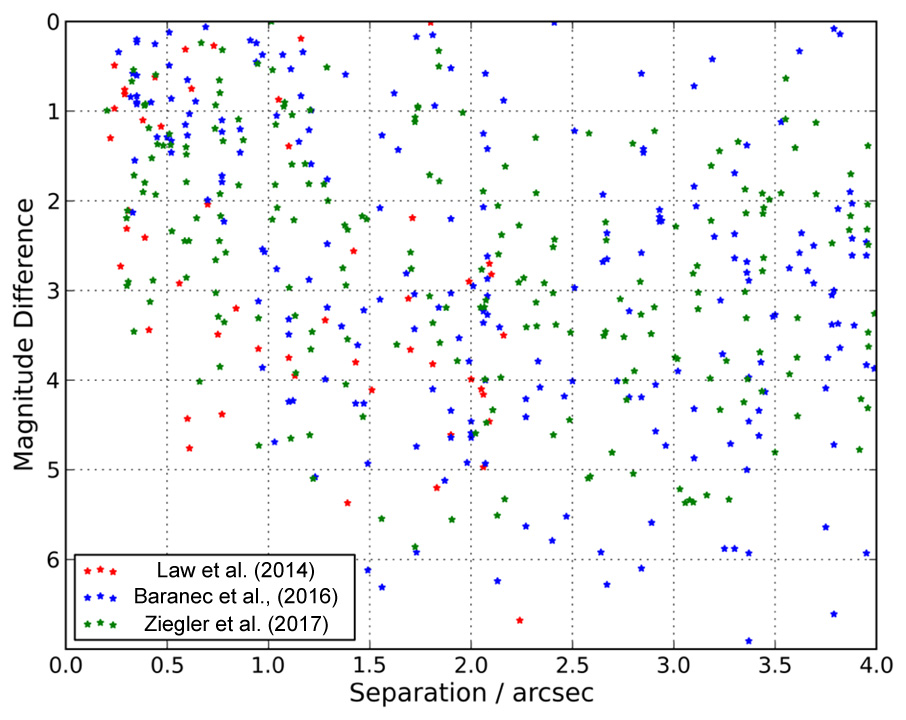}}
\caption{Separations and magnitude differences of the detected companions in the full Robo-AO Kepler Planetary Candidate Survey.}
\label{ra_companions}
\end{figure}

We operated Robo-AO in its automated mode for $\sim$180 nights at Palomar, spanning June 2012 to June 2015. On nights with no weather or environmental losses, Robo-AO typically completed 200 -- 250 observations, each 90s to 120s of total integration time (sufficient to get to the photon noise floor set by the uncorrected seeing halo in the first few arc seconds of separation). During its time at Palomar, the system completed $\sim$19,000 observations (see Fig. \ref{sky_map}), comprising several AO surveys with the most numerous observations ever performed \citep{Riddle14B}. This includes known stars within 25 pc observable from Palomar and nearly all of \textit{Kepler's} candidate exoplanet host stars. For the latter, we discovered 479 stars within 4$''$ of 3313 \textit{Kepler} host stars, yielding a nearby star fraction of 14.5$\pm0.8\%$ (\citealt{Ziegler16}; Fig. \ref{ra_companions}). Approximately half of these nearby stars are within 2$''$, a separation range where only high-angular resolution surveys are able to accurately measure the properties of the companion stars.\looseness=-1

\section{Future directions}

In November 2015, we relocated the Robo-AO system to the Kitt Peak 2.1-m telescope for a 3-year deployment \citep{JC17}. The system was augmented with a near-infrared avalanche photodiode array camera, enabling simultaneous imaging with the visible camera. While visible and infrared tip-tilt correction were demonstrated previously, these new modes will be wrapped into the current robotic observing software.

Future Robo-AO systems will be deployed at other modest sized telescope apertures, including the University of Hawai`i 2.2-m telescope at Maunakea where the superior seeing will naturally enhance the achieved image quality of observations - and where we can take advantage of more recent component advancements, e.g., MEMS DMs with a greater number of actuators and EMCCD wavefront sensor detectors. In addition to imaging, autonomous AO can play a crucial role in enhancing the sensitivity of low spectral resolution instruments for rapidly characterizing transients and other time-domain phenomena \citep{RTS}. This will be particularly critical for the larger telescope apertures necessary to fully characterize the wealth of new discoveries in the era of the Large Synoptic Survey Telescope.

\section{Acknowledgments}
 
C.B. acknowledges support from the Alfred P. Sloan Foundation. The Robo-AO system was developed by collaborating partner institutions, the California Institute of Technology and the Inter-University Centre for Astronomy and Astrophysics, and with the support of the National Science Foundation under Grant Nos. AST-0906060, AST-0960343, and AST-1207891, the Mt. Cuba Astronomical Foundation and by a gift from Samuel Oschin. Some of the research presented is supported by the NASA Exoplanets Research Program, grant $\#$NNX 15AC91G.  Robo-AO at Kitt Peak was supported by a grant from the Murty family, who feels very happy to have added a small value to this important project, as well as grants from the Mt. Cuba Astronomical Foundation and the John Templeton Foundation.

\bibliographystyle{ws-rv-van}
\bibliography{ws-rv-sample}

\end{document}